\documentclass[sigconf, nonacm=true]{acmart}

\usepackage[ruled,vlined]{algorithm2e}
\usepackage{graphicx}

\newcommand{\E}{\mathbb{E}}

\newcommand{\bX}{\mathbf{X}}

\newcommand{\bZ}{\mathbf{Z}}
\newcommand{\be}{\mathbf{e}}
\newcommand{\bx}{\mathbf{x}}
\newcommand{\bw}{\mathbf{w}}
\newcommand{\bbeta}{\boldsymbol{\beta}}
\newcommand{\betahat}{\boldsymbol{\widehat{\beta}}}

\newcommand{\Sigmahat}{\boldsymbol{\widehat{\Sigma}}}

\renewcommand{\l}{\left}
\renewcommand{\r}{\right}

\begin{document}

\title{Randomized Controlled Trials without Data Retention}
\author{Winston Chou}
\email{winston_chou@apple.com}
\affiliation{
	\institution{Apple}
	\city{}
	\state{}
	\country{}
}

\begin{abstract}
    Amidst rising appreciation for privacy and data usage rights, researchers have increasingly acknowledged the principle of \emph{data minimization}, which holds that the accessibility, collection, and retention of subjects' data should be kept to the bare amount needed to answer focused research questions.  Applying this principle to randomized controlled trials (RCTs), this paper presents algorithms for making accurate inferences from RCTs under stringent data retention and anonymization policies.  In particular, we show how to use recursive algorithms to construct running estimates of treatment effects in RCTs, which allow individualized records to be deleted or anonymized shortly after collection.  Devoting special attention to non-i.i.d. data, we further show how to draw robust inferences from RCTs by combining recursive algorithms with bootstrap and federated strategies.
\end{abstract}

\keywords{randomized controlled trials, data minimization, A/B testing}

\maketitle

\section{Introduction}

Randomized controlled trials (RCTs) are the gold standard for causal evidence in medicine, technology, and numerous other fields.  In a typical RCT, researchers collect data on $n$ units, which are randomly allocated into treatment and control groups.  At the conclusion of the experiment, when a pre-specified time or sample size threshold is met, statistics are computed over these groups and compared for evidence of a causal effect.

Conventional analytic procedures for RCTs require researchers to store all $n$ individual records until the end of the trial.  Yet, such retention is not strictly necessary and carries privacy risks for experiment participants.  To temper such risks, this paper presents methods for computing precise statistics from RCT data that do not require individual records to be stored for the duration of the trial.  In particular, we outline methods that comply with strict data retention policies that guarantee that subjects' individual-level data will be deleted or anonymized shortly after collection.  Such policies, which are increasingly ubiquitous in digital platforms and enshrined in documents like the GDPR, embody the principle of \emph{data minimization}, which holds that the accessibility, collection, and retention of subjects' data should be kept to the bare amount needed to address targeted research questions \citep{pfitzmann2010terminology}.

The algorithms make use of several families of statistical methods.  First, we use recursive estimation to keep running tallies of parameter estimates, which are updated with each new observation.  This means that raw, individualized records can be discarded or anonymized immediately after being incorporated into the collective estimate.  Second, to improve statistical precision using pre-experiment data, we extend these methods to linear regression models of RCT data \citep{deng2013improving}.  As such methods allow researchers to detect causal effects more quickly and with less data, they constitute yet another application of the data minimization principle.  Lastly, to address the commonplace challenge of non-i.i.d. records -- which create special problems for data deletion and anonymization practices -- we show how recursive estimation can be combined with bootstrap and federated methods for robust inference.


\section{Experimentation Without Data Retention}
\label{sec:problem}

The motivating problem for this paper is to estimate the average effect of a treatment $d$ on an outcome $y$ in an RCT whilst minimizing the time that individual-level data is stored for analysis.  Raw observations consist of individual records $(y_i, \bx_i)$, $i \in 1, \ldots, n$, where $y_i$ is unit $i$'s measurement on $y$ and $\bx_i$ is a column vector of features.  $\bx_i$ includes a binary treatment indicator $d_i$, which equals 1 if record $i$ is randomly assigned to the treatment group and 0 otherwise.  We assume that records arrive continuously throughout the RCT, which is may be longer in duration than mandated data retention periods.  We further assume that the non-treatment features in $\bx_i$ are unaffected by the treatment (for example, they may be measured prior to the RCT).\footnote{We also impose standard regularity conditions on $\bx$, e.g., the features should not be linear combinations of each other.}

Our quantity of interest is the population average treatment effect (PATE):
\begin{equation}
    \tau = \E[y_i^1 - y_i^0],
\end{equation}
where $y_i^1$ is the $i$-th unit's potential outcome under treatment, $y_i^0$ is the same unit's potential outcome under control, and the expectation is taken over the population represented by the RCT sample \citep{imbens2015causal}.\footnote{We omit discussion of sampling issues from this paper.}

In the absence of data retention limits, estimating the PATE is trivial: simply subtracting the mean of $y$ in the control group from the mean of $y$ in the treatment group yields an unbiased estimate.  Moreover, we can leverage the additional features in $\bx$ to achieve more precise estimates of the PATE using linear regression \citep[][]{deng2013improving}, among other methods.  Yet, conventional implementations of these estimators assume that each individual record can be kept until all records have been observed; our aim is to show how these estimators can be implemented without this assumption.

This paper contributes to the literature on digital experimentation by showing how recursive estimation can be used to overcome common challenges in RCTs -- mainly, protection of research subjects' privacy through swift data deletion, but also variance reduction, heteroscedasticity, and clustering.  A related paper \citep{coey:coir:haim:liou:2020} similarly contemplates how to analyze cluster-randomized experiments with limited data retention, demonstrating that the pseudo-random sampling technique in \citet{bakshy2013uncertainty} can solve this problem (see also Section~\ref{sec:bootstrap} of this paper).  We extend their methodology to linear regression; identify a potential privacy risk; and propose a federated alternative to eliminate this risk.

\section{Proposed Methodology}
\label{sec:method}


 The basic idea behind our proposed algorithms is to keep a running tally of estimates, which is updated whenever a new record is observed.  Only the most recent set of estimates, along with the new observation, are needed to update the tally, meaning that individual records can be discarded immediately after being incorporated into the collective estimate.  To illustrate, we present the simple example of computing the mean of a sequence of measurements, denoted $z_1, z_2, \ldots, z_n$.  At any given $t \in \{1, \ldots, n\}$, the arithmetic mean of the $z_i$'s observed up to that point can be computed as:
 \begin{equation}
     \label{eqn:mean}
     \overline{z}_t = \frac{1}{t} \sum_{i=1}^t z_i.
 \end{equation}
 Note that we do not need to retain each individual value of $z_i$, $i < t$ to estimate $\overline{z}_t$.  Instead, we need only to keep track of the number of observations and sum of $z_i$'s prior to $t$.  Then, once the $t$-th observation is made, we can add $z_t$ to the running sum and divide by $t$ to compute the mean.

 In fact, Equation~\ref{eqn:mean} can be written \emph{recursively} to express this update in even fewer steps:
 \begin{eqnarray}
     \label{eqn:recursive_mean}
     \overline{z}_t &=& \overline{z}_{t-1} + \frac{1}{t}\l(z_t - \overline{z}_{t-1}\r).
 \end{eqnarray}
 As Equation~\ref{eqn:recursive_mean} makes clear, retaining the individual records $z_1, \ldots, z_{t-1}$ is unnecessary to compute $\overline{z}_t$.  The information contained in these observations is wholly encapsulated in $\overline{z}_{t-1}$, which was itself computed by updating $\overline{z}_{t-2}$ with $z_{t-1}$, etc.  As such, we can discard each record immediately after incorporating it into the running tally.\footnote{Equation~\ref{eqn:recursive_mean} can be generalized to batch updates.  Let $t^\prime$ denote the total record count after observing the batch and $\Delta = \sum_{i=t+1}^{t^\prime} z_i$ denote the batch sum.  Then $\overline{z}_{t^\prime} = \overline{z}_t + \frac{1}{t^\prime}[\Delta - (t^\prime - t) \overline{z}_t)$].}



 From Equation~\ref{eqn:recursive_mean}, it is clear that we can obtain an unbiased estimate of the PATE from an RCT without needing to store any individual records for the duration of the trial.  Let $z_i = \frac{d_i y_i}{\pi_1} - \frac{(1 - d_i) y_i}{1 - \pi_1}$, where $\pi_1 = \E[d_i]$ is the probability of being assigned to the treatment group.  Upon observing all $n$ records, the recursive estimate of $\mathbb{E}[z_i]$ is equivalent to the arithmetic mean, which in turn equals:
 \begin{eqnarray}
     \overline{z}_n &=& \frac{1}{n} \sum_{i=1}^n \l(\frac{d_i y_i}{\pi_1} - \frac{(1 - d_i) y_i}{1 - \pi_1}\r) \\
     &=& \frac{1}{n} \l(\sum_{i=1}^n \frac{d_i y_i}{\pi_1} - \sum_{i=1}^n \frac{(1 - d_i) y_i}{1 - \pi_1}\r).
 \end{eqnarray}
 Taking expectations, we have that:
 \begin{eqnarray}
     \E[\overline{z}_n] &=& \frac{1}{n} (n \E[y_i | d_i = 1] - n \E[y_i | d_i = 0]) \\
     &=& \E[y_i | d_i = 1] - \E[y_i | d_i = 0] \\
     &=& \mathbb{E}[y_i^1 - y_i^0],
 \end{eqnarray}
 where the last equality is implied by random assignment.  

\subsection{Recursive Linear Regression for RCTs}
\label{sec:rls}

We begin by considering the linear regression model for RCTs:
\begin{equation}
    \label{eqn:ols}
    y_i = \bx_i^\top\bbeta + e_i,
\end{equation}
where, as before, $\bx_i = [1, d_i, \ldots]^\top$ is a $k$-length feature vector that contains the treatment indicator $d_i$ as well as a constant offset; $\bbeta$ is the column vector of linear regression coefficients $[\beta_1, \ldots, \beta_k]^\top$ whose second component $\beta_2$ corresponds to the marginal treatment effect (i.e., the coefficient on $d_i$); and $e_i$ is a stochastic error term having mean 0.  Note that Equation~\ref{eqn:ols} implies that the conditional mean of the control potential outcome, $\E[y_i^0 | \bx_i]$, is a linear function of the non-treatment features in $\bx$.

The benefit of fitting this model to RCT data is that, when the non-treatment features in $\bx$ are strongly correlated with the outcome $y$, the statistical variance of the estimated treatment effect is considerably reduced \citep{deng2013improving}.  As a result, less data is needed to detect the treatment effect and/or measure it precisely.  Moreover, when the treatment is randomly assigned (as in an RCT), the least squares estimate of $\beta_2$ in this model converges to the PATE in large samples, even if the relationship between the outcome and other features is not truly linear \citep{imbens2015causal, lin2013agnostic}.

In classical linear regression, the error terms $e_i$ are assumed to be i.i.d. normal with variance $\sigma^2$.  Under this assumption, the sampling variance of $\betahat$ is consistently estimated by:
\begin{equation}
    \Sigmahat_{IID} = \l(\sum_{i=1}^n \bx_i\bx_i^\top\r)^{-1} \widehat{\sigma^2},
    \label{eqn:olsvar}
\end{equation}
where:
\begin{equation}
    \widehat{\sigma^2} = \frac{1}{n - k - 1} \sum_{i=1}^n \hat{e}_i^2 = \frac{1}{n - k - 1} \sum_{i=1}^n (\bx_i^\top\betahat - y_i)^2,
\end{equation}
which can be used to compute asymptotically valid $t$-statistics and confidence intervals for $\betahat$.

To fit $\betahat$ and $\Sigmahat_{IID}$ with minimal data retention, we propose the method of recursive least squares \citep{harvey1990econometric}.  Letting $\betahat_t$ denote the OLS estimate of $\boldsymbol{\beta}$ after observing $t$ records and $(y_{t+1}, \bx_{t+1})$ denote a new record, the updated estimate $\betahat_{t+1}$ is given by:
\begin{equation}
    \label{eqn:beta}
    \betahat_{t+1} = \betahat_t + (\bX_{t+1}^\top\bX_{t+1})^{-1} \bx_{t+1}(y_{t+1} - \bx_{t+1}^\top\betahat_t),
\end{equation}
where $\bX_m$ denotes the $m \times k$ matrix of features corresponding to the initial $m \le n$ observations.  


Note that $(\bX_t^\top\bX_t)^{-1}$ can also be updated directly:
\begin{equation}
    \label{eqn:cov}
    (\bX_{t+1}^\top\bX_{t+1})^{-1} = (\bX_t^\top\bX_t)^{-1} - \frac{(\bX_t^\top\bX_t)^{-1} \bx_{t+1}\bx_{t+1}^\top (\bX_t^\top\bX_t)^{-1}}{1 + \bx_{t+1}^\top (\bX_t^\top\bX_t)^{-1} \bx_{t+1}},
\end{equation}
which simplifies the task of computing Equation~\ref{eqn:beta}.

The final step is to compute the residual standard error $\widehat{\sigma^2}$, which constitutes the other piece of $\Sigmahat_{IID}$.  Letting $S_t$ denote the sum of squared residuals after $t$ observations:
\begin{equation}
    S_t = \sum_{i=1}^t (\bx_i^\top\betahat - y_i)^2,
\end{equation}
$S_{t+1}$ is computed as:
\begin{equation}
    \label{eqn:ssr}
    S_{t+1} = S_t + \frac{(\bx_{t+1}^\top\betahat_t - y_{t+1})^2}{1 + \bx_{t+1}^\top(\bX_t^\top\bX_t)^{-1}\bx_{t+1}}.
\end{equation}

Putting these pieces together yields the recursive least squares (\texttt{RLS}) algorithm (Algorithm~\ref{algo:rls}).

\begin{algorithm}
\SetAlgoLined
\DontPrintSemicolon
 Initialize estimates of $\betahat$, $(\bX^\top\bX)^{-1}$, and $S$, for example using the first $m \ge 0$ records.\;
 \For{each record $i = m + 1, \ldots, n$}{
  $S \gets S + \frac{(\bx_i^\top\betahat - y_i)^2}{1 + \bx_i^\top(\bX^\top\bX)^{-1}\bx_i}$\;
  $(\bX^\top\bX)^{-1} \gets (\bX^\top\bX)^{-1} - \frac{(\bX^\top\bX)^{-1} \bx_i\bx_i^\top (\bX^\top\bX)^{-1}}{1 + \bx_i^\top (\bX^\top\bX)^{-1} \bx_i}$\;
  $\betahat \gets \betahat + (\bX^\top\bX)^{-1} \bx_i(y_i - \bx_i^\top\betahat)$\;
 }
 \caption{Recursive least squares (\texttt{RLS}) \citep{harvey1990econometric}}
 \label{algo:rls}
\end{algorithm}

\subsection{Bootstrap Inference for Recursive Estimation}

Inference in RCTs does not only consist of point estimates of the treatment effect; it also consists of quantifying the uncertainty of such estimates.  Above, we showed how to estimate the sampling variance of $\betahat_n$ using normal approximations.  However, the reliability of such approximations will depend on the sampling distribution of the error terms $e_i$, which can depart significantly from normality in practice.

A more robust approach is to use bootstrap methods, which estimate the sampling variability by resampling the original records with replacement \citep{efron2016computer}.\footnote{We can also estimate heteroscedasticity robust standard errors (HRSEs) when the $e_i$ are independent but have different variances \citep{wooldridge2010econometric}.  HRSEs are commonly applied to RCTs, since treatment effect heterogeneity implies heteroscedasticity.  We show how to estimate HRSEs recursively in the Appendix.}  However, the conventional bootstrap assumes that we have kept the individual-level records for analysis, which is exactly what our recursive algorithms seek to avoid.

Recursive estimation works well with the online bootstrap \citep{owen2012bootstrapping}, which does not require all of the data to be collected in advance \citep{chamandy2012estimating}.  In the online bootstrap, each new record is assigned a random vector of $B$ weights, $\bw_i = [w_i^{(1)}, \ldots, w_i^{(B)}]^\top$, where each weight is drawn independently from a distribution with unit mean and variance.  The resulting set of $B$ statistics -- with the $i$-th unit contributing to the $B$-th statistic in proportion to $w_i^{(b)}$ -- is then used as an approximation to the sampling distribution of that statistic.

To tailor the online bootstrap to recursive estimation, we first initialize $B$ bootstrap sample statistics $\boldsymbol{\overline{z}}_0 = [\overline{z}_0^{(1)}, \ldots, \overline{z}_0^{(B)}]^\top$.  Once the $t$-th record arrives, a $B$-length vector of weights, denoted $\bw_t$, is sampled from a Poisson(1) distribution.  The record then updates all $B$ statistics in proportion to the components of $\bw_t$.  Once all $n$ records are observed, the vector $\boldsymbol{\overline{z}}_n = [\overline{z}_n^{(1)}, \ldots, \overline{z}_n^{(B)}]^\top$ is treated as the bootstrap distribution for $\overline{z}$.

We provide a simple numerical example based on the PATE.  Suppose we are running an RCT with probability of treatment $\pi_1 = 0.50$.  The first record is a treated observation having $y_1 = 2$.  In the conventional bootstrap methodology, we would store $y_1$ individually until all $n$ records have arrived, then resample those records $B$ times, such that first record appears (say) once in the first resample, twice in the second resample, and not at all in the $B$-th resample.  These counts can be represented as the vector of weights $\bw_1 = [1, 2, \ldots, 0]^\top$.

In the online bootstrap, rather than resample the full dataset, the weight vectors are drawn from a probabilistic distribution as records arrive.\footnote{See \citet{chamandy2012estimating} for theoretical justification of this approach for the Poisson(1) distribution.}  Accordingly, suppose that the same weight vector $\bw_1$ were randomly sampled instead.  Upon observing $y_1 = 2$, $d_1 = 1$, and $\bw_1 = [1, 2, \ldots, 0]^\top$, the updated value of $\boldsymbol{\overline{z}}_1$ is given by $[\frac{1 \times z_1}{1}, \frac{2 \times z_1}{2}, \ldots, 0] = [4, 4, \ldots, 0]^\top$, where $z_1 = \frac{d_1 y_1}{\pi_1} - \frac{(1 - d_1) y_1}{1 - \pi_1} = 4$.

Having computed $\boldsymbol{\overline{z}}_1$, we can discard the raw information $(y_1, d_1)$ and instead update $\boldsymbol{\overline{z}}_1$ using the following recurrence relation for the (weighted) mean:
\begin{eqnarray}
    \label{eqn:weighted_recursive_mean}
    \overline{z}_t &=& \begin{cases}
    	\overline{z}_{t-1} & \text{if $w_t = 0$} \\
    	\overline{z}_{t-1} + \frac{w_t}{n_t}\l(z_t - \overline{z}_{t-1}\r) & \text{otherwise},
    \end{cases}
\end{eqnarray}
where we have redefined $n_t$ as $\sum_{i=1}^t w_t$, the sum of weights up to the $t$-th observation.  For example, if we next observed a control record with $y_2 = 1$ and $\bw_2 = [2, 1, \ldots, 1]^\top$, the updated value of $\boldsymbol{\overline{z}}_2$ would be $[4 - \frac{12}{3}, 4 - \frac{6}{3}, \ldots, 0 - 2]^\top = [0, 2, \ldots, -2]^\top$.  Repeating this procedure (Algorithm~\ref{algo:bootstrap}) for all $n$ observations yields $B$ estimates -- constituting the bootstrap distribution of $\overline{z}_n$ -- of the PATE.

\begin{algorithm}
\SetAlgoLined
\DontPrintSemicolon
\SetKwFor{ForParallel}{for}{in parallel do}{}
 Initialize $B$ bootstrap sample means $\overline{z}^{(1)}, \ldots, \overline{z}^{(B)}$ and pseudocounts $n^{(1)}, \ldots, n^{(B)}$ at zero\;
 \For{each record $i = 1, \ldots, n$}{
  $z_i \gets \frac{d_i y_i}{\pi_1} - \frac{(1 - d_i) y_i}{\pi_0}$\;
  Sample $B$ weights $w_i^{(1)}, \ldots, w_i^{(B)}$ from Poisson(1)\;
  \ForParallel{each bootstrap sample $b = 1, \ldots, B$}{
  $n^{(b)} \gets n^{(b)} + w_i^{(b)}$\;
  $\overline{z}^{(b)} \gets \overline{z}^{(b)}$ if $w_i^{(b)} = 0$ else $\overline{z}^{(b)} + \frac{w_{i}^{(b)}}{n^{(b)}}\l(z_i - \overline{z}^{(b)}\r)$ \;
  }
 }
 \caption{Online bootstrap for the PATE}
 \label{algo:bootstrap}
\end{algorithm}

It is possible to extent Algorithm~\ref{algo:bootstrap} to linear regression by replacing the inner loop with weighted \texttt{RLS}.\footnote{The standalone weighted \texttt{RLS} algorithm is presented in the Appendix.}  However, this raises the tricky question of how to initialize the recursion: with very few observations, $\bX^\top\bX$ is not invertible, and so the starting values of $\betahat$ will be unidentified.  As a workaround, in Algorithms~\ref{algo:rls} and~\ref{algo:bootstrap-rls}, we suggest fitting starting values using the first $m$ records and ordinary batch methods.  The starting values can then used to initialize all $B$ recursions.  Naturally, the resulting bootstrap distribution will understate the variance, but decreasingly so as $n$ grows relative to $m$.

\begin{algorithm}
\SetAlgoLined
\DontPrintSemicolon
\SetKwFor{ForParallel}{for}{in parallel do}{}
Estimate $\betahat$ and $\bZ \equiv (\bX^\top\bX)^{-1}$ using the first $m$ records\;
Initialize $B$ bootstrap coefficients $\betahat^{(1)}, \ldots, \betahat^{(B)}$ and matrices $\bZ^{(1)}, \ldots, \bZ^{(B)}$ at $\betahat$ and $\bZ$, respectively\;
 \For{each record $i = m+1, \ldots, n$}{
  Sample $B$ weights $w_i^{(1)}, \ldots, w_i^{(B)}$ from Poisson(1)\;
  	\ForParallel{each bootstrap sample $b = 1, \ldots, B$}{
  	$\bZ^{(b)} \gets \bZ^{(b)} - \frac{w_i^{(b)}\bZ^{(b)} \bx_i\bx_i^\top \bZ^{(b)}}{1 + w_i^{(b)} \bx_i^\top \bZ^{(b)} \bx_i}$\;
  	$\betahat^{(b)} \gets \betahat^{(b)} + \bZ^{(b)} w_i^{(b)}\bx_i(y_i - \bx_i^\top\betahat^{(b)})$\;
  }
 }
 \caption{Online bootstrap for weighted \textsf{RLS}}
 \label{algo:bootstrap-rls}
\end{algorithm}

\subsection{Bootstrap Inference with Dependent Data}
\label{sec:bootstrap}

It is well-known that the standard errors given by Equation~\ref{eqn:olsvar} are inappropriate for cluster-randomized trials (and dependent data generally)~\citep{abadie2017should}.  While there are a number of ways to adjust variance estimates for clustering, one of the most robust is the cluster bootstrap.  In the typical implementation of this method, clusters rather than individual records are resampled with replacement.  This requires that we are able to stitch together records from the same higher-level unit (e.g., users in user-randomized trials) over time, meaning that the records cannot be anonymized, let alone deleted, until all records have arrived.

To relax this requirement, we can use the online bootstrap algorithm with deterministic sampling \citep{bakshy2013uncertainty, coey:coir:haim:liou:2020}.  In the online bootstrap, each record is assigned a random vector of $B$ weights, $\bw_i = [w_i^{(1)}, \ldots, w_i^{(B)}]^\top$, where each weight is drawn independently from a distribution with unit mean and variance.  The resulting set of $B$ statistics -- with the $i$-th unit contributing to the $B$-th statistic in proportion to $w_i^{(b)}$ -- is then used as an approximation to the sampling distribution of that statistic.  A ``trick'' for extending this method to clustered data is to seed the random weight generator with the cluster identifier for the $i$-th record, denoted by $j[i]$, such that every record belonging to $j[i]$ will receive an identical vector of weights $\bw_{j[i]}$.  This emulates the conventional cluster bootstrap, as each record will appear in the same frequency in each bootstrap resample as all of the other records in the same cluster.

When combined with recursive estimation, this yields a straightforward algorithm for estimating sampling variances with clustered observations;  Algorithm~\ref{algo:clusterboot} exposits in the case of the PATE.  Note that the only difference from Algorithm~\ref{algo:bootstrap} is that the weights are no longer i.i.d. across records but generated using $j[i]$ as a seed.

\begin{algorithm}
\SetAlgoLined
\DontPrintSemicolon
\SetKwFor{ForParallel}{for}{in parallel do}{}
 Initialize at zero $B$ bootstrap sample means $\overline{z}^{(1)}, \ldots, \overline{z}^{(B)}$ and pseudocounts $n^{(1)}, \ldots, n^{(B)}$\;
 \For{each record $i = 1, \ldots, n$}{
  Sample $B$ weights $w_{j[i]}^{(1)}, \ldots, w_{j[i]}^{(B)}$, using $j[i]$ as a seed\;
  $z_i \gets \frac{d_i y_i}{\pi_1} - \frac{(1 - d_i) y_i}{\pi_0}$\;
  \ForParallel{each bootstrap sample $b = 1, \ldots, B$}{
  $n^{(b)} \gets n^{(b)} + w_i^{(b)}$\;
  $\overline{z}^{(b)} \gets \overline{z}^{(b)}$ if $w_i^{(b)} = 0$ else $\overline{z}^{(b)} + \frac{w_{i}^{(b)}}{n^{(b)}}\l(z_i - \overline{z}^{(b)}\r)$ \;
  }
 }
 \caption{Online cluster bootstrap for the PATE}
 \label{algo:clusterboot}
\end{algorithm}

\subsection{Federated Computation of Cluster-Robust Standard Errors}
\label{sec:federated}
A potential flaw of Algorithm~\ref{algo:clusterboot} is that, when records are only anonymized and not deleted, the bootstrap weights $\bw_{j[i]}$ can be used to associate records from the same higher-level unit -- that is, they serve as an implicit ``fingerprint.''  


To be sure, this risk is not unique: All methods of adjusting for intracluster correlation assume \emph{some} ability to associate records from the same cluster.  This association is critical to drawing valid inferences from RCTs, but greatly complicates data deletion and anonymization practices.

To avoid this complication, we also propose a \emph{federated} alternative to cluster-robust inference.  In federated computation, research subjects (commonly called clients) store and perform computations on their individual records \citep{konevcny2016federated}.  These independent computations are collected by the server -- possibly without client identifiers -- and immediately aggregated into collective statistics.  Consequently, individualized records need neither be communicated to nor retained on the server.  Of course, federated computation implies the existence of clients that are capable of storing and performing computations on their own data (e.g., mobile phones).  It may also be subject to communication costs that compel us to limit the size and number of transmissions to a fixed length and frequency \citep{konevcny2016federated}.

Under communication constraints, an appealing direction is to use analytical \emph{cluster-robust standard errors} (CRSEs) for inference \citep{liang1986longitudinal}.\footnote{CSREs are appropriate whenever the treatment is assigned to higher-level units, but raw observations are collected at a finer granularity \citep{abadie2017should}.  Precisely speaking, CSREs are appropriate when the structural mean model in Equation~\ref{eqn:ols} is correct, but the error terms $e_i$ are correlated within each cluster.}  As the name implies, CRSEs adjust $\Sigmahat_{IID}$, the estimated variance-covariance matrix of $\betahat$, to account for intracluster correlation of records from the same unit.  Usefully, each unit's contribution to the CRSEs can be computed independently and communicated to the server as a $k$-length vector, itself an aggregation of that unit's granular records.

Our proposed implementation of CRSEs is as follows.  As in Section~\ref{sec:rls}, records are streamed to the server during the RCT to generate recursive estimates of $\betahat$ and $(\bX^\top\bX)^{-1}$.  These can be quickly discarded and/or anonymized (or, in contrast with the online cluster bootstrap, collected without subject identifiers) on the \emph{server}.  Each client retains its own records for the duration of the experiment.  At the conclusion of the RCT, the server pushes the final estimates of the linear regression coefficients $\betahat_n$ to the clients.  The clients then compute their independent contributions to the variance of $\betahat_n$, given below.  These contributions are collected and averaged on the server to produce the final estimate $\Sigmahat_n$. 

This procedure emulates the following batch estimator of the variance-covariance matrix of $\betahat$:
\begin{equation}
    \Sigmahat_{CL} = (\bX^\top\bX)^{-1} \bX^\top \be \be^\top \bX (\bX^\top\bX)^{-1},
\end{equation}
where $\be = [\hat{e}_1, \ldots, \hat{e}_n]^\top$ is the vector of $n$ linear regression residuals.\footnote{We assume (WLOG, since reordering the data is trivial in the batch processing case) that records are ordered by cluster.}  $\Sigmahat_{CL}$ is often called the ``sandwich estimator'' because it sandwiches a ``meat'' matrix, $\bX^\top \be \be^\top \bX$, between two ``bread'' matrices $(\bX^\top\bX)^{-1}$ \citep{wooldridge2010econometric}.  Under clustering, $\be \be^\top$ is a block-diagonal matrix, which means the meat matrix can be rewritten as
\begin{equation}
    \label{eqn:cluster_meat}
    \bX^\top \be \be^\top \bX = \sum_j \bX_j^\top \be_j \be_j^\top \bX_j,
\end{equation}
where $\bX_j$ is the $n_j \times k$ matrix consisting of the $j$-th cluster's feature vectors and $\be_j$ is the column vector of its regression residuals.  Note that each unit's contribution to $\Sigmahat_{CL}$ (that is, $\bX_j^\top e_j e_j^\top \bX_j$) can be computed independently and anonymously aggregated to form $\bX^\top \be \be^\top \bX$, which can in turn be combined with recursive estimates of $(\bX^\top\bX)^{-1}$ to compute $\Sigmahat_{CL}$.

\begin{figure}[ht]
    \centering
    \includegraphics[width=0.8\linewidth]{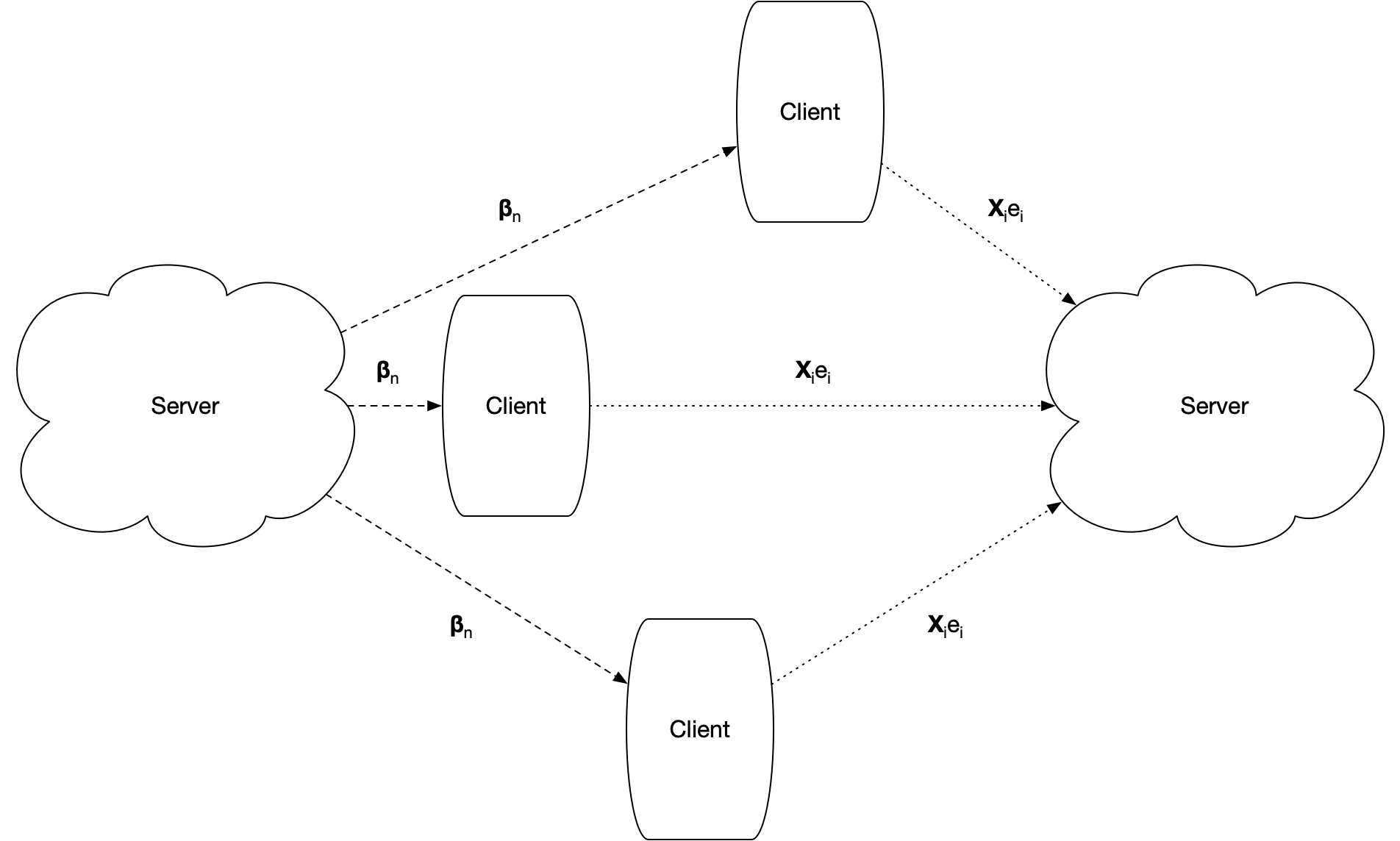}
    \caption{Federated computation of cluster-robust standard errors. Once the server pushes the final estimates of $\bbeta$ to the clients, each client independently computes its contribution to $\Sigmahat_{SW}$, then passes these, possibly anonymously, to the server for aggregation.  All messages consist of $k$ numbers.}
    \label{fig:federated}
\end{figure}

Figure~\ref{fig:federated} illustrates.  In the first stage, the server pushes the final estimates of $\bbeta$ to the clients.  Each client then computes $\bX_j^\top\be_j = \sum_{i=1}^{n_j} \bx_i (\bx_i^\top \betahat_n - y_i)$ and communicates it to the server, possibly without identifiers.  Lastly, the server computes the outer products of these vectors and sums the resulting $k \times k$ matrices to compute $\bX^\top \be \be^\top \bX$ and $\Sigmahat_{CL}$.

This computation bears some resemblance to federated learning for linear regression \citep{mcmahan2017communication}.  For instance, in both algorithms, the client updates consist of the squared error loss gradient computed on the client's own data.  However, because our focus is on estimating the variance of the fitted model parameters $\betahat_n$ and not how these parameters are fitted, we require just one round of back-and-forth communication between the server and clients.\footnote{In the special case where the treatment assignment is the only independent variable in the regression formula, the CRSE can be computed based on the delta method:
\begin{equation}
	Var\left(\sum_{i=1}^n y_i / n \right) \approx \frac{1}{J \overline{n_j}} \l( \sigma_s^2 + \frac{\overline{s_j}^2}{\overline{n_j}^2} \sigma_n^2 - 2 \frac{\overline{s_j}}{\overline{n_j}} \sigma_{s,n} \r),
	\label{eqn:delta}
\end{equation}
where $n_j$ and $s_j$ are the size and sum of $y$ for the $j$-th cluster, respectively; $\overline{n_j}$ and $\overline{s_j}$ are their averages over clusters; $\sigma_n^2$ and $\sigma_s^2$ are their variances; and $\sigma_{s, n}$ are their covariance \citep{deng2021equivalence}.  No back-and-forth communication is needed to compute~\ref{eqn:delta}; each client just needs to communicate its values of $n_j$ and $s_j$ (and treatment assignment).}

\begin{algorithm}
\SetAlgoLined
\DontPrintSemicolon
\SetKwInOut{Input}{input}
\SetKwInOut{Output}{output}
\SetKwFor{ForParallel}{for}{in parallel do}{}
 Push $\betahat$ to clients

 \ForParallel{each client $j = 1, \ldots, J$}{
    \ForParallel{each record $i = 1, \ldots, n_j$}{
        $e_i \gets \bx_i^\top\betahat - y_i$\;
    }
    $\mathbf{u_j} \gets \sum_i^{n_j} \bx_i e_i$\;
    return $\mathbf{u_j}$ to server\;
 }
 $\Sigmahat_{SW} \gets (\sum_{i=1}^n \bx_i\bx_i^\top)^{-1}\sum_{j=1}^J \mathbf{u_j} \mathbf{u_j}^\top (\sum_{i=1}^n \bx_i\bx_i^\top)^{-1}$
 \caption{Federated computation of cluster-robust standard errors}
 \label{algo:federated}
\end{algorithm}

\section{Conclusion}
\label{sec:conclusion}
We have outlined methods for recursively computing precise statistics from RCTs.  Because recursive estimation eliminates the need to store raw, individualized records for analysis, it is an attractive option for working with data that is sensitive in nature and/or subject to stringent retention or anonymization policies.

The primary motivation for our methods is to enable the swift deletion -- even the noncollection -- of RCT data.  There are, to be sure, fundamental tensions between this goal and open science principles like data transparency and replicability.  Yet, there are also points of alignment; e.g., committing to recursive estimation implies pre-specifying an analysis plan and greatly limits the scope for $p$-hacking and post-hoc specification search \citep{munafo2017manifesto}.  Future research should explore methods of enabling transparency and replicability that also respect fundamental privacy and data usage rights.
%

\bibliographystyle{ACM-Reference-Format}
\bibliography{bib}

\end{document}